\documentclass[aps,pra,twocolumn,showpacs,amsmath,amssymb,preprintnumbers,superscriptaddress,10pt]{revtex4-2}
\usepackage{graphicx}
\usepackage{float}
\usepackage{subfigure}
\usepackage{dcolumn}
\usepackage{bm}
\usepackage{hyperref}
\usepackage{siunitx}
\usepackage{amsmath,amssymb}
\usepackage{marvosym}
\usepackage{braket}
\usepackage{booktabs}
\usepackage{appendix}
\usepackage{ragged2e}
\hypersetup{colorlinks=true, citecolor=blue, urlcolor=blue, linkcolor=blue}

\begin{document}
	

\title{Deep Anomaly Detection for Active Attacks on the Receiver in \protect\\ Quantum Key Distribution}

\author{Junxuan Liu}
\affiliation{College of Computer Science and Technology, National University of Defense Technology, Changsha 410073, People's Republic of China}

\author{Bingcheng Huang}
\affiliation{College of Computer Science and Technology, National University of Defense Technology, Changsha 410073, People's Republic of China}

\author{Jialei Su}
\affiliation{College of Computer Science and Technology, National University of Defense Technology, Changsha 410073, People's Republic of China}

\author{Qingquan Peng}
\affiliation{College of Computer Science and Technology, National University of Defense Technology, Changsha 410073, People's Republic of China}

\author{Anqi~Huang}
\email{angelhuang.hn@gmail.com}
\affiliation{College of Computer Science and Technology, National University of Defense Technology, Changsha 410073, People's Republic of China}
	
\begin{abstract}
    Traditional countermeasures against attacks targeting the receiver in quantum key distribution (QKD) systems often suffer from poor compatibility with deployed infrastructure, the risk of introducing new vulnerabilities, and limited applicability to specific types of active attacks. In this work, we propose an anomaly detection (AD) model based on one-class machine learning to address active attacks targeting the receiver. By constructing a dataset from the QKD system’s operational states, the AD model learns the characteristics of normal behavior under secure conditions. When an active attack occurs, the system's state deviates from the learned normal patterns and is identified as anomalous by the model. Experimental results show that the AD model achieves an area under the curve (AUC) exceeding 99\%, effectively safeguarding the receiver of the QKD system. Compared to traditional approaches, our model can be deployed with minimal cost in existing QKD networks without requiring additional optical or electrical components, thus avoiding the introduction of new side channels. Furthermore, unlike multi-class machine learning algorithms, our approach does not rely on prior knowledge of specific attack types and is potentially able to detect unknown active attacks. These advantages—generality, ease of deployment, low cost, and high accuracy—make our model a practical and effective tool for protecting the receiver of QKD systems against active attacks.

\end{abstract}
\maketitle

\section{Introduction} \label{sec:Introduction}

Quantum key distribution (QKD), based on the quantum mechanism, such as quantum no-cloning theorem and Heisenberg’s uncertainty principle, enables the information-theoretical security of distributing random symmetric keys~\cite{bennett1984,RevModPhys.74.145,RevModPhys.81.1301,RevModPhys.92.025002}. Compared to classical cryptographic algorithms that rely on computational complexity, QKD offers significant advantages in terms of security, which have attracted widespread attention. However, in practical implementations, the physical devices used in QKD often deviate from the idealized theoretical models. In recent years, extensive research into these discrepancies has revealed various side channels, demonstrating that an eavesdropper (Eve) can exploit such vulnerabilities to compromise the practical security of QKD systems~\cite{Lydersen2010,Xu_2010,Wiechers_2011,Gerhardt2011,PhysRevLett.107.110501,PhysRevLett.112.070503,7571108,PhysRevA.98.012330,PhysRevApplied.10.064062,PhysRevApplied.12.064043,PhysRevApplied.13.034017,e24020260,Chaiwongkhot2022,PhysRevA.106.033713,PRXQuantum.3.040307,PhysRevApplied.19.014048,PhysRevApplied.21.014026,Xing:24,Liu:24,Peng2025}.

To address side channels arising from imperfections in physical devices, researchers have explored multiple approaches to protect QKD systems from quantum hacking. One such approach is the development of new QKD protocols. For example, the measurement-device-independent QKD (MDI~QKD) protocol eliminates security dependencies on imperfect measurement devices~\cite{PhysRevLett.108.130503}. While MDI~QKD offers higher security compared to the decoy-state BB84 protocol~\cite{PhysRevLett.94.230503,PhysRevLett.94.230504,PhysRevA.72.012326}, it also introduces significantly higher implementation complexity. Importantly, since most practical QKD systems still rely on mature prepare-and-measure protocols~\cite{Liao2017,PhysRevLett.120.030501,Chen2021} that remain vulnerable to quantum attacks, it is crucial to develop low-complexity countermeasures that can enhance their security without changing existing infrastructures. 

One strategy involves adding physical components to the system as a “patching” approach to counter specific attacks~\cite{PhysRevX.5.031030,PhysRevA.94.030302,PRXQuantum.3.040307}. While this method is relatively easy to implement, it may inadvertently introduce new side channels, especially for active attacks targeting the receiver~\cite{Wu:20}, thereby potentially compromising the overall security of the QKD system. Another strategy for enhancing security without altering existing infrastructures focuses on refining theoretical security models.  This involves analyzing Eve’s capabilities under specific attacks and estimating the corresponding secure key rates ~\cite{PhysRevX.5.031030,Tamaki_2016,Zapatero2021securityofquantum,PhysRevApplied.18.044069,Xing:24,Liu:24}. By incorporating the characteristics of known attacks into the theoretical framework, these models aim to quantify potential key leakage. For certain passive attacks where Eve exploits inherent imperfections of physical devices—such as correlations, misalignments in modulation, or detection backflash~\cite{PereiraSciAdv,Lu:22,su2025quantifying}—to extract key information, it is suggested to adopt theoretical security models that incorporate these imperfections. However, the QKD system usually fails to predict the active attacks and is hard to consider them in the security proof in advance. This situation frequently happens for the unexpected active attacks on the receiver side of a QKD system, which cannot be simply protected by isolation components.

These limitations highlight the need for a more general defense capability against a wide range of attack strategies, while still minimizing changes to existing infrastructures. To address this challenge, this work proposes an intelligent countermeasure for QKD systems based on machine learning—an anomaly detection (AD). In particular, we employ Deep Support Vector Data Description (Deep SVDD)~\cite{pmlr-v80-ruff18a}, a one-class classification algorithm, to achieve real-time monitoring of QKD systems and ensure their operation in a secure environment. Our AD model is trained under unsupervised conditions, which greatly simplifies dataset preparation. For instance, in our experiment, the parameters from the calibration and post-processing stages can be directly extracted during the secure operation of the QKD system to construct the training dataset. We constructed a test set by combining QKD system parameters obtained under both secure conditions and under attacks (specifically, the calibration attack~\cite{PhysRevLett.107.110501} and the muted attack~\cite{su2025mutedattackhighspeedquantum}) in a 1:1 ratio. Experimental results demonstrate that Deep SVDD achieves an area under the receiver operating characteristic curve (AUC) exceeding 99\%. This indicates that the AD model can effectively detect anomalies using only QKD system parameters.

With the advantages of machine learning, our AD model not only reduces implementation costs but also avoids introducing new side channels. Meanwhile, the data construction, training, and testing of the machine learning model can be performed in parallel with key distribution, ensuring that the operation of QKD system is not affected~\cite{Lu:19,PhysRevA.105.042411,Xu2024,refId0}. However, unlike multi-class classification approaches~\cite{PhysRevA.105.042411,Xu2024,refId0}, which attempt to distinguish between specific attack types, our one-class AD model is designed to flag any behavior that causes deviations in system parameters, regardless of the attack's nature. This enables the model could be able to detect even previously unknown or unmodeled active attack strategies, enhancing the robustness and security performance of QKD system. Importantly, our experimental results show that the anomaly detection performance of the AD model depends on the richness of the training dataset. By feeding the model with more diverse and representative training data, the underlying neural network can learn a broader range of system behavior patterns, thereby enhancing its anomaly detection capability. Overall, our approach offers advantages in generality, simplicity, low cost, and high accuracy, paving the way for more robust and scalable QKD system security in future deployments.

The paper is structured as follows. Section~\ref{sec:Method} introduces the construction of AD model, including the neural network architecture and dataset preparation. Section~\ref{sec:Result} presents the training process and testing results of the model. In Sec.~\ref{sec:Discussion}, we compare the AD model with other countermeasures and discuss the AD model’s application scenarios and limitations. Finally, Sec.~\ref{sec:Conclusion} concludes the study.

\section{Anomaly Detection Model Design and Dataset Construction} \label{sec:Method}

\begin{figure}[htbp]
\centering
\includegraphics[width=\linewidth]{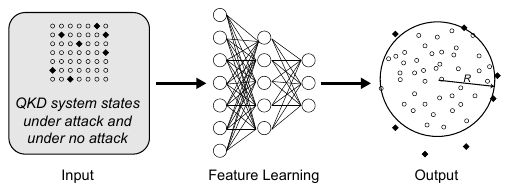}
\caption{\justifying{The conceptual diagram of the AD model. Hollow circles represent the QKD system's states under secure conditions, while solid squares denote states generated under attack. These state parameters are extracted to form the dataset for anomaly detection. During training, the neural network learns only from normal data to construct a hypersphere that encloses their distribution. Any anomalous data falling outside this hypersphere is identified as abnormal, thus enabling effective anomaly detection.}}
\label{Fig:concept}
\end{figure}

In this section, we present the design of the AD model and the construction of the corresponding dataset. The dataset is built by capturing the operational states of the QKD system. Parameters collected under secure conditions are treated as normal data, while those recorded during attacks are labeled as anomalies. Importantly, the neural network is trained exclusively on the normal dataset to learn representative features—without relying on any anomalous samples. These features are then used by the Deep SVDD algorithm to construct a hypersphere that encloses the distribution of normal data, such that anomalous inputs fall outside the hypersphere and can be effectively detected. Built upon neural networks, Deep SVDD leverages end-to-end representation learning and can be extended to handle more complex tasks, albeit at the cost of increased computational overhead. Owing to this capability, it is particularly well suited for high-dimensional and nonlinear data, making it an appropriate choice for our application scenario. The overall concept of the AD model is illustrated in Fig.~\ref{Fig:concept}.

    \subsection{Deep Support Vector Data Description and Neural Network Modeling}

    To achieve general and real-time monitoring of QKD systems, we adopt the AD model based on Deep SVDD~\cite{pmlr-v80-ruff18a}, a deep one-class classification model. The core idea of Deep SVDD is to extend traditional SVDD~\cite{Tax2004} by integrating it with deep neural networks, thereby enabling the learning of expressive feature representations from structured or high-dimensional inputs. Unlike multi-class classification approaches that require labeled data from various attack types, Deep SVDD requires only data collected under normal operating conditions for training, without any prior knowledge of attack behaviors. This makes it especially suitable for QKD scenarios, where anomalous data (e.g., from new or rare attacks) may be unavailable during training. During deployment, the model learns a compact feature space for normal data and identifies any significant deviations as anomalies, regardless of the specific nature of the underlying attack. This generalization ability allows the model to detect a broad range of potential threats—including previously unknown attacks—by solely analyzing the operational parameters of a running QKD system. A detailed explanation of both SVDD and Deep SVDD is provided in Appendix~\ref{SVDD&DSVDD}.

    In our AD model, we consider that only a small number of parameters can be extracted from the QKD system. Therefore, we adopt the multilayer perceptron (MLP) architecture as the backbone network for Deep SVDD, due to its simplicity, low computational cost, and effectiveness in small-scale feature learning. To fit our dataset, we adjust the input layer to match the dimensionality of the QKD parameter vector and configure hidden layers to extract representative features from normal data. These features are then used to construct a hypersphere enclosing the distribution of normal samples. Leveraging the representation learning capability of MLP, the customized Deep SVDD model transforms normal QKD data into compact, informative latent representations. Through forward propagation, each network layer encodes progressively higher-level abstractions of the QKD parameter space. The model is trained using the Deep SVDD objective, which minimizes the volume of a hypersphere enclosing the latent features of normal data—ensuring that the network captures the typical behavior of secure QKD operation.

    Training is performed via backpropagation, iteratively updating the network weights to reduce the hypersphere loss. As the model converges, it forms a reliable boundary around the normal data distribution. At inference time, any input whose representation falls outside this hypersphere is flagged as anomalous, indicating a potential deviation from secure operation. This design enables effective anomaly detection without relying on labeled attack samples or prior knowledge of the attack method, making it well-suited for real-world QKD systems.

    \subsection{Dataset Construction for Anomaly Detection}

    In this subsection, we describe the construction of the datasets used to train and evaluate the AD model. The guiding principle is that the parameters recorded during the operation of the QKD system should faithfully reflect its internal state and behavior. In this context, two categories of parameters are collected. One category consists of configuration parameters, such as detector gate position signals and the voltage values of polarization controllers. The other includes consists of non-configuration parameters such as the timestamps of single-photon detector (SPD) responses. By capturing these operational characteristics, the dataset enables the AD model to effectively detect anomalies introduced by malicious attacks. From the perspective of Eve, we consider how different attack strategies affect the QKD system and accordingly propose two methods for constructing datasets based on the nature of these attacks. In the first type of active attack, Eve manipulates the system by modifying its configuration parameters. For instance, Eve may shift the timing of detector gate signals through the calibration attack~\cite{PhysRevLett.107.110501}, change the bias voltage of intensity modulators via the induced-photorefraction attack~\cite{PhysRevApplied.19.054052,Lu:23,PhysRevApplied.20.044013}. To capture the effects of such intrusions, the QKD system can be equipped with monitoring interfaces that continuously record these configuration parameters in real time, forming a dataset that reflects system behavior under various conditions. In the second type of active attack, Eve may not alter any system configuration parameters directly. This type of attacks, such as, the muted attack~\cite{su2025mutedattackhighspeedquantum} and the superlinear attack~\cite{PhysRevA.84.032320}, manipulate the SPD responses without changing the system's internal settings. In these cases, we focus on the SPD's output behavior—particularly the timing of detection events—and use the associated timestamps to construct a dataset that captures anomalies induced by this class of attacks.

    For the construction of datasets based on QKD system configuration parameters, we extract a wide range of relevant information from as many stages of the QKD operation as possible. As a concrete example, we consider a discrete-variable QKD system employing polarization encoding. The QKD process can be broadly divided into three stages: the calibration stage, the raw key exchange stage, and the post-processing stage. During the calibration stage, the receiver (Bob) continuously adjusts the timing of the gate signals for his SPDs to synchronize with the photon pulses sent by the transmitter (Alice). Simultaneously, Bob tunes the internal polarization controller (PC) to ensure proper alignment with Alice’s polarization encoding. Both the gate timing values and the PC settings are recorded as part of the AD dataset during this stage. In the raw key exchange stage, Alice and Bob perform quantum signal transmission and reception to generate raw keys. No system configuration parameters are typically adjusted during this phase. In the post-processing stage, Alice and Bob perform a series of statistical evaluations. These include the number of sifted keys, the ratio of signal detections to decoy state emissions, the ratio of signal to decoy state detections, the detection efficiencies corresponding to signal, decoy, and vacuum states, as well as the quantum bit error rates (QBERs) for each polarization basis and the overall QBER. During error correction and privacy amplification, additional parameters—such as privacy amplification factors—are also generated. Since these statistical quantities objectively reflect the operational state of the QKD system, they are likewise included in the AD dataset.
    
    To construct the dataset for attacks that do not alter system configuration parameters, we extract the timestamps of SPD responses as input data for anomaly detection. Specifically, in our experimental setup, the SPDs are connected to a time-to-digital converter (TDC), which records the arrival time of each detection event. The TDC operates with a 100 ns cycle, capturing the precise time at which each SPD response occurs. As a result, all data in this dataset represent detection timestamps, with values ranging from 0 to 100 ns. Unlike the previous dataset based on system configuration and statistical parameters, this dataset is directly constructed from individual SPD counts, allowing the feature dimension to scale with the number of detection events. In our experiments, we generated datasets with feature dimensions corresponding to 100, 225, and 400 detection events, respectively. These datasets were then used to train and test the anomaly detection model. Based on the performance of the AD model, we further identified the optimal feature dimension for this type of dataset.

    It is worth emphasizing that the data of parameters is simply concatenated in a certain sequence to form a vector with no preprocessing. This makes our data collection process simple and efficient. For normal data, both datasets are generated by operating the QKD system in a secure environment. We then select the calibration attack~\cite{PhysRevLett.107.110501} and the muted attack~\cite{su2025mutedattackhighspeedquantum} to generate anomalous samples because these attacks currently cannot be effectively mitigated through adding physical components or refining theoretical security model. Moreover, experimental results have shown that both attacks are capable of obtaining sifted keys from the QKD system, thereby severely compromising its security. Considering the nature of these two attacks, the anomalous data for the first dataset are collected under the calibration attack, while those for the second dataset are collected under the muted attack. During the training of the AD model, only normal data are used, whereas the testing phase involves datasets composed of normal and anomalous samples in a 1:1 ratio.

\section{Model Training and Performance Evaluation} \label{sec:Result}

As described in Sec.~\ref{sec:Method}, we constructed two types of datasets. Accordingly, the same AD model was trained and evaluated separately on each dataset. The training was performed using identical hyperparameter settings, as summarized in Table~\ref{tab: hyperparameter}. To evaluate the model’s performance in detecting anomalies, we used the AUC as the evaluation metric~\cite{ERFANI2016121,pmlr-v80-ruff18a}. A detailed explanation of the AUC is provided in Appendix~\ref{AUC}.

\begin{table}[htbp]
\caption{\label{tab: hyperparameter}%
Training hyperparameters for the AD model.}
\begin{ruledtabular}
\begin{tabular}{cccc}
Optimizer & Learning Rate & Batch Size & Epochs \\
\hline
Adam & 0.0001 & 128 & 150 \\

\end{tabular}
\end{ruledtabular}
\end{table}

\begin{figure}[htbp]
    \centering
    \includegraphics[width=\linewidth]{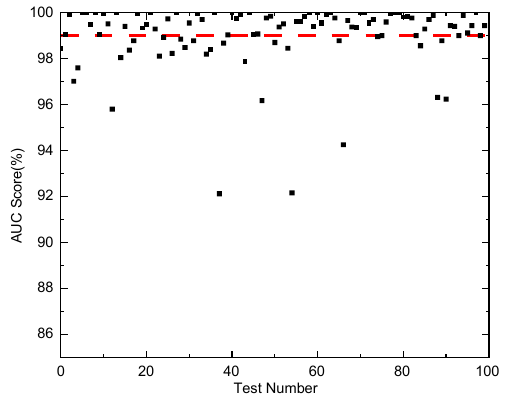}
    \caption{\justifying{AUC results obtained from 100 independent tests using the configuration parameters dataset. The dataset is constructed from the parameters of the calibration and post-processing stages, while the anomalous data were generated under the calibration attack. The black dots represent the actual AUC values obtained from each test, and the red dashed line indicates the average AUC across all 100 runs.}}
    \label{Fig:test_calibration}
    \end{figure}

We first train and evaluate the AD model using a dataset consisting of configuration parameters extracted from the calibration and post-processing stages of the QKD system. To mitigate the influence of experimental randomness, the trained AD model is independently tested on different test sets 100 times. The resulting AUC values are shown in Fig.~\ref{Fig:test_calibration}. Across these 100 tests, the average AUC reached 99.03\%, with the minimum value still exceeding 92\%~\footnote{It is worth noting that in repeated experiments a few test runs yielded AUC values noticeably below the average one. This phenomenon can be attributed to the sensitivity of Deep SVDD to randomness in training (e.g., initialization, dropout, and optimizer stochasticity), which may cause convergence to different local minima. }. These results demonstrate that the AD model, trained solely on parameters collected under normal conditions, can effectively learn the characteristics of legitimate system behavior. When subjected to attacks that alter system parameters, such as the calibration attack, the model achieves an anomaly detection rate exceeding 99\%, thereby enabling reliable identification of Eve's presence.

    \begin{figure}[htbp]
    \centering
    \includegraphics[width=\linewidth]{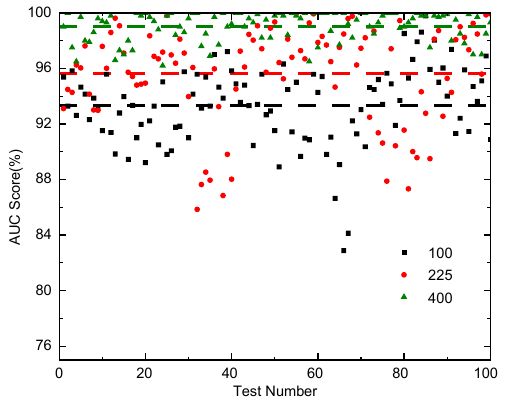}
    \caption{\justifying{AUC results obtained from 100 independent tests using the dataset of SPD response timestamps. The dataset is constructed from the timestamps corresponding to SPD counts, with the anomalous data generated under the muted attack. The black square dots represent the AUC values obtained by testing the AD model trained on a dataset constructed using 100 SPD counts, while the black dashed line indicates the average AUC across the 100 tests. Similarly, the red circular dots and the green triangular dots (along with their corresponding dashed lines) represent the AUC values (and their averages across the 100 tests) when the dataset is constructed using 225 and 400 SPD counts, respectively.}}
    \label{Fig:test_muted}
    \end{figure}

As discussed in Sec.~\ref{sec:Method}, we constructed the dataset using the timestamps of SPD counts and evaluated the anomaly detection capability of the AD model against the muted attack. Similarly, to ensure the robustness of the results, the trained model was evaluated 100 times, each with a different testing set. The resulting AUC values are shown in Fig.~\ref{Fig:test_muted} and Table~\ref{tab: Statistics}. It is demonstrated that, the anomaly detection performance of the AD model improves with the number of SPD counts used in the dataset. When the detection events reach 400, the average AUC value over 100 independent tests is 99.03\%, indicating a high detection accuracy. Moreover, the variance is only 1.09, demonstrating the stability of the AD model.
\begin{table}[htbp]
\caption{\label{tab: Statistics}%
The mean and variance of test AUC for feature dimensions of 100, 225, and 400.}
\begin{ruledtabular}
\begin{tabular}{cccc}
 & \multicolumn{3}{c}{\text{Feature dimension}} \\
\text{Statistics} & 100 & 225 & 400 \\
\hline
Mean(\%) & 93.35 & 95.65 & 99.03 \\
Variance & 8.07 & 12.45 & 1.09 \\

\end{tabular}
\end{ruledtabular}
\end{table}

\begin{figure}[htbp]
    \centering
    \includegraphics[width=\linewidth]{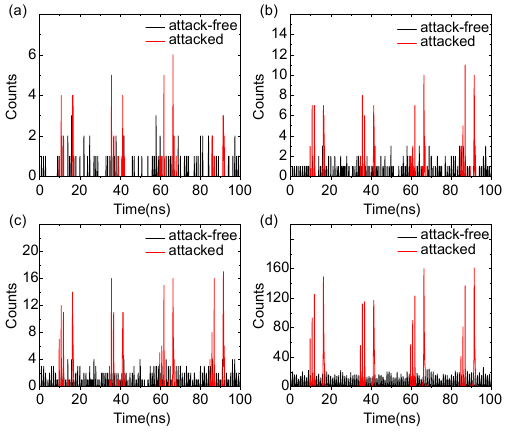}
    \caption{\justifying{Timestamp distributions under different numbers of detection events. (a)–(d) show the histogram of SPD timestamps within a \SI{100}{ns} cycle, using a bin size of \SI{0.1}{ns}, for total detection counts of $100$, $225$, $400$, and $4000$, respectively. The black histograms represent data collected under secure conditions, while the red histograms correspond to data collected under muted attack.}}
    \label{Fig:ana_muted}
    \end{figure}

To further investigate why a larger data sample size enhances anomaly detection performance, we analyze the structure of the dataset derived from TDC timestamps. Each \SI{100}{ns} cycle is divided into bins of \SI{0.1}{ns}, and statistical histograms of overlapping detection events are generated for sample sizes of 100, 225, and 400 counts, as shown in Fig.~\ref{Fig:ana_muted}. For comparison, we also present a histogram based on 4000 counts. From Fig.~\ref{Fig:ana_muted}, we observe that under attack-free conditions, SPD responses exhibit an approximately uniform distribution, whereas under the muted attack, the responses become more concentrated in specific regions as explained in Ref.~\cite{su2025mutedattackhighspeedquantum}. When the number of detection events is small [e.g., Fig.~\ref{Fig:ana_muted}(a)], the distribution difference between normal and anomalous data is not particularly pronounced. However, as the sample size increases, the discrepancy in the temporal distribution becomes increasingly evident. This divergence allows the AD model to extract more discriminative features, thereby improving its ability to detect anomalies. In this context, the feature dimension—determined by the number of detection events—is positively correlated with detection performance. Nevertheless, as the AD model already achieves an AUC above 99\% with low variance at 400 detection events, further increasing the feature dimension would significantly raise the complexity of data collection and the computational cost of model training and inference. Therefore, we adopt 400 detection events as the input size for our AD model, balancing detection accuracy with resource efficiency.

Through training and testing experiments on the AD model using two different datasets, it is demonstrated that the AD model can achieve excellent anomaly detection performance by training solely on datasets generated under secure conditions in the QKD system. The average AUC values obtained over multiple tests exceed 99\%, indicating that the AD model has effectively learned the characteristics of normal QKD system. When Eve launches an active attack, the AD model can promptly and accurately detect the anomaly, thereby helping to guard the practical security of the QKD system.

\section{Discussion} \label{sec:Discussion}

In this section, we present a comparative analysis between the proposed AD model and conventional methods, as well as a comparison with strategies based on multi-class classification approaches. At the same time, we also discuss the application scenarios and limitations of the AD model.

Ensuring the practical security of existing QKD infrastructures is essential. Unlike approaches that require the design and deployment of new QKD protocols, the AD method proposed here operates entirely within existing systems and requires no additional optical or electrical hardware. By extracting operational parameters directly from deployed QKD systems, our model enables effective detection of anomalous behavior, helping to maintain system security in real time. Furthermore, since our method leverages internally generated QKD data rather than relying on specific implementation details, it provides strong portability and general applicability across a wide range of QKD architectures.

Adding physical devices to defend against attacks is a common “patching” strategy in QKD system design. Such countermeasures are typically tailored to specific attacks. However, while potentially effective against known threats, this approach increases the physical complexity and cost of QKD system implementation. Moreover, the introduction of additional hardware components may itself create new, unforeseen side channels, thereby compromising the practical security of the system. For instance, a countermeasure designed to detect the intensity of injected light—intended to prevent detector blinding attacks—was experimentally shown to be ineffective~\cite{Wu:20}. By contrast, our proposed AD model avoids these drawbacks. It enables legitimate users to monitor the operational status of the QKD system in real time for signs of abnormal behavior, without requiring any additional optical or electrical hardware. As a result, it does not introduce new potential vulnerabilities that could be exploited by Eve. Furthermore, since the AD module is deployed within the trusted environment (e.g., inside the room of the legitimate parties), and in QKD, Eve is not permitted to access this area physically, the security of the AD model can be reasonably assumed.

Another line of countermeasure involves refining the theoretical security models of QKD systems. This approach focuses on analyzing Eve's ability to compromise the key through specific attacks and estimating the corresponding secure key rate. However, this approach has inherent limitations. It relies on prior knowledge of specific attack mechanisms and explicitly incorporates the resulting information leakage into the security analysis. As a result, it offers no protection against unknown or yet-to-be-modeled attacks. For certain types of active attacks targeting the receiver—such as the calibration attack and muted attack used in this work to evaluate the AD model—there currently exist no effective theoretical modeling tools. In contrast, our proposed AD model does not rely on a priori modeling of specific active attack strategies. Instead, it identifies anomalies by detecting deviations in QKD system parameters, regardless of the underlying attack mechanism. This makes the AD model not only effective against known and modeled attacks, but also potentially capable of detecting previously unknown or uncharacterized threats.

Based on machine learning, our method adopts a one-class anomaly detection approach rather than a multi-class classification algorithm~\cite{PhysRevA.105.042411,Xu2024,refId0}. This choice is motivated by the principle that the primary objective of intelligent countermeasures in QKD systems is to detect the presence of attacks, rather than to classify specific attack types. While multi-class models can identify known attacks in the training dataset, they are ineffective against previously unknown attacks. In contrast, our AD model is formulated as an unsupervised learning method, meaning that it only requires parameters from normal QKD operation during training, without any labels indicating attack types. Using a multilayer perceptron, the AD model captures the features of normal parameters, and any deviation caused by attacks can be detected, ensuring secure operation. This design endows the model with generalization ability, allowing it to detect even unknown attack strategies. Moreover, the unsupervised nature of our AD framework greatly simplifies data preparation. In practical experiments, raw data such as time-stamped detector counts can be directly used to construct the dataset without manual labeling. The detection capability of the model is closely linked to the richness and diversity of the training data. A more comprehensive dataset allows the neural network to learn a broader and more nuanced representation of the system’s normal behavior, enhancing its anomaly detection performance. Therefore, careful design of a dataset that fully captures the characteristics of QKD system parameters is essential for the robust deployment of the AD model.

In practical QKD implementations, the quantum channel is inevitably subject to environmental disturbances, which may introduce fluctuations in both noise levels and transmittance. Since the channel is regarded as untrusted, our AD model does not rely on any assumptions about its behavior. Thus, we do not distinguish whether the observed noise originates from an eavesdropper’s intervention or from environmental fluctuations. In either case, its effect is ultimately reflected in the final key rate. Therefore, the logic of key generation under noise is that the AD model first determines whether the current QKD system is under active attacks. If the system is deemed secure, the key is generated based on the specific value of QBER or excess noise. If the AD model identifies a potential attack as anomalous, the corresponding key is discarded.

Although our experiments demonstrate that the AD model can efficiently and accurately identify active attacks, it is important to note that in practice the hypersphere constructed by Deep SVDD may not perfectly separate normal and anomalous data in all cases. Instead, the method relies on empirical risk minimization. We acknowledge that if future white-box attacks specifically target machine learning methods, our AD model might fail to recognize them. Nevertheless, Deep SVDD has been extensively validated as one of the most effective anomaly detection frameworks in practice and has been applied in various domains, demonstrating strong potential in complex, high-dimensional, and dynamic scenarios~\cite{pmlr-v80-ruff18a,ruff2019deep,ZHANG20211}. Thus, we adopted Deep SVDD in designing our AD model for QKD systems, contributing meaningful progress toward enabling anomaly detection as a defense mechanism in QKD. Our experiments confirm that the AD model can effectively detect abnormal parameters under realistic QKD conditions, thereby verifying its feasibility from an empirical perspective. We also hope that future work will explore theoretical guarantees or multi-center extensions of Deep SVDD to further enhance robustness.

\section{Conclusion} \label{sec:Conclusion}

This paper proposes a machine learning-based AD method to achieve real-time monitoring of QKD system parameters. By building an AD model based on Deep SVDD, we guard that the QKD system operates under secure conditions. The datasets used for model training and evaluation are constructed from parameters recorded during the QKD system's operational process, including system configuration parameters and detector response timestamps. The training set exclusively consists of data collected when the QKD system operates securely. For testing, we generate a balanced dataset (1:1 ratio) combining data from both secure and attacked conditions. Two classical active attack strategies — the calibration attack and the muted attack — are used to produce anomalous data in the test set.  Testing results show that it can achieve an AUC of up to 99.03\%, indicating a high capability to detect the presence of active attacks targeting the receiver. Our AD approach offers generality, simplicity, low cost, and high accuracy, making it a promising solution for ensuring the practical security of QKD systems and providing valuable insights for the design of future QKD architectures.

~

\begin{acknowledgments}
    This work was funded by the National Natural Science Foundation of China (Grant No. 62371459), the Innovation Program for Quantum Science and Technology (2021ZD0300704), and the Postdoctoral Fellowship Program of CPSF (Grant No. GZC20252817). 
\end{acknowledgments}

\begin{appendix}
\appendix

\section{Support Vector Data Description and Deep Support Vector Data Description} \label{SVDD&DSVDD}

Our AD model is based on SVDD and Deep SVDD proposed in Ref.~\cite{Tax2004} and Ref.~\cite{pmlr-v80-ruff18a}. To make our paper self-contained, we present the main results from Ref.~\cite{Tax2004} and Ref.~\cite{pmlr-v80-ruff18a} in this section.

Traditional SVDD is a kernel-based method that maps input data from the original feature space into a high-dimensional feature space via a feature mapping function~\cite{Tax2004}. In this space, SVDD aims to find the smallest hypersphere, with center $\mathbf{c} \in \mathcal{F}_k$ and radius $R > 0$, that encloses most of the normal data points. Data points that lie outside the hypersphere are considered anomalies. The objective function is formulated as:
\begin{widetext}
    \begin{equation}
    \min_{R, \mathbf{c}, \boldsymbol{\xi}} \; R^2 + \frac{1}{\nu n} \sum_{i} \xi_i
    ~~~~~~~\text{s.t.} \quad \| \phi_k(\mathbf{x}_i) - \mathbf{c} \|_{\mathcal{F}_k}^2 \leq R^2 + \xi_i, \quad \xi_i \geq 0, \quad \forall i
    \label{SVDD}
    \end{equation}  
\end{widetext}
    Here, the slack variables $\xi_i > 0$ allow for a soft boundary, and the hyperparameter $\nu \in (0,1]$ controls the trade-off between the volume of the hypersphere and the penalty for outliers. Points satisfying $\| \phi_k(\mathbf{x}_i) - \mathbf{c} \|_{\mathcal{F}_k}^2 > R^2$ are considered anomalous.

    However, traditional SVDD heavily relies on the choice of kernel functions, making it difficult to capture the intrinsic structure and features of complex nonlinear data. In addition, the storage and computational complexity of the kernel matrix is $\mathcal{O}(n^2)$, which limits its scalability to large datasets. Moreover, the hypersphere's center and radius typically need to be predefined, which reduces the model's robustness to shifts in the data distribution~\cite{NIPS2007_013a006f,5419028}. The Deep SVDD model differs from traditional SVDD by replacing the kernel function mapping with a deep neural network~\cite{pmlr-v80-ruff18a}. It directly learns nonlinear features in the original input space, enabling automatic learning of hierarchical and abstract representations of the data. This forms an end-to-end deep one-class classification framework that can more effectively distinguish between normal and abnormal data.

    In the following, we present the main principles of Deep SVDD as introduced in Ref.~\cite{pmlr-v80-ruff18a}. For some input space $\mathcal{X} \subset \mathbb{R}^d$ and output space $\mathcal{F} \subset \mathbb{R}^p$, let $\phi(\cdot; \mathbf{W}) : \mathcal{X} \rightarrow \mathcal{F}$ be a neural network with $L \in \mathbb{N}$ hidden layers and a set of weights $\mathbf{W} = \{ \mathbf{W}^1, \dots, \mathbf{W}^L \}$, where $\mathbf{W}^\ell$ are the weights of layer $\ell \in \{1, \dots, L\}$. That is, $\phi(\mathbf{x}; \mathbf{W}) \in \mathcal{F}$ is the feature representation of $\mathbf{x} \in \mathcal{X}$ given by the network $\phi$ with parameters $\mathbf{W}$. The aim of Deep SVDD is to jointly learn the network parameters $\mathbf{W}$ while minimizing the volume of a data-enclosing hypersphere in the output space $\mathcal{F}$, characterized by a radius $R > 0$ and center $\mathbf{c} \in \mathcal{F}$, which is assumed to be fixed for now. Given some training data $\mathcal{D}_n = \{ \mathbf{x}_1, \dots, \mathbf{x}_n \} \subset \mathcal{X}$, the \textit{soft-boundary} Deep SVDD objective is:
\begin{widetext}
    \begin{equation}
    \min_{R, \mathbf{W}} \; R^2 + \frac{1}{\nu n} \sum_{i=1}^{n} \max\left\{ 0, \left\| \phi(\mathbf{x}_i; \mathbf{W}) - \mathbf{c} \right\|^2 - R^2 \right\} + \frac{\lambda}{2} \sum_{\ell=1}^{L} \left\| \mathbf{W}^{\ell} \right\|_F^2
    \label{DeepSVDD}
    \end{equation}
    where the second term contains the loss function for each data point
    \begin{equation}
    \mathcal{L}(\mathbf{x}_i) = \max\left\{ 0, \; \| \phi(\mathbf{x}_i; \mathbf{W}) - \mathbf{c} \|^2 - R^2 \right\}.
    \end{equation}
\end{widetext}
    In Eq.(\ref{DeepSVDD}), the first term minimizes the volume of the hypersphere, while the second term penalizes data points that lie outside the hypersphere after passing through the network. The hyperparameter $\nu \in (0,1]$ controls the trade-off between the hypersphere volume and the violation of its boundary, allowing some points to be mapped outside the hypersphere. The final term is a weight decay regularizer for the network parameters. This objective function is suitable for training scenarios where a small number of normal samples may lie outside the hypersphere due to feature fluctuations, making it more applicable to datasets that may contain slight anomalies.

\section{The Area Under the Receiver Operating Characteristic Curve and Its Calculation} \label{AUC}

In classification tasks within machine learning, predictive models typically output numerical values representing prediction scores or estimated probabilities that a sample belongs to a particular class. For one-class classification, a threshold must be predetermined to make binary decisions: if a sample's score exceeds this threshold, it is classified as normal; otherwise, it is labeled as anomalous. It is evident that the choice of threshold directly affects the classification results and, consequently, alters the composition of the confusion matrix, which records the relationship between predicted labels and ground truth.

The receiver operating characteristic curve demonstrates the performance of a classifier by systematically depicting the relationship between the true positive rate (TPR) and the false positive rate (FPR) under different thresholds. The formulas for TPR and FPR are given as follows:
\begin{equation}
\text{TPR} = \frac{\text{TP}}{\text{TP} + \text{FN}}
\end{equation}
\begin{equation}
\text{FPR} = \frac{\text{FP}}{\text{FP} + \text{TN}}
\end{equation}
The meanings of TP, FN, FP, and TN are shown in Table~\ref{tab:confusion-matrix}.

\begin{table}[H]
\centering
\caption{Confusion matrix of classification results.}
\label{tab:confusion-matrix}
\begin{ruledtabular}
\begin{tabular}{ccc}
& \multicolumn{2}{c}{\text{Predicted}} \\
\text{Actual} & Positive & Negative \\
\hline
Positive & True Positive (TP) & False Negative (FN) \\
Negative & False Positive (FP) & True Negative (TN) \\
\end{tabular}
\end{ruledtabular}
\end{table}

The AUC represents the area under the receiver operating characteristic curve and provides a quantitative measure of the overall performance of the classifier. It is defined as:
\begin{equation}
\text{AUC} = \int_0^1 \text{TPR}(\text{FPR}) \, d\text{FPR}
\end{equation}
The AUC is fundamentally a probabilistic measure. It represents the likelihood that, when randomly selecting one normal and one anomalous sample, the classifier assigns a higher score to the normal sample than to the anomalous one. From this definition, a higher AUC value implies that the classifier is more capable of distinguishing between normal and anomalous samples, thereby indicating better classification performance.

The AUC is particularly suitable for imbalanced datasets, as it is insensitive to the ratio of normal to anomalous samples. Moreover, the AUC provides an intuitive and interpretable metric to evaluate the classifier's discriminative power between the two classes. An AUC value of 1 indicates that there exists at least one threshold at which the classifier can perfectly separate all normal and anomalous samples—an ideal case that is rarely achieved in real-world scenarios. When the AUC lies between 0.5 and 1, the classifier performs better than random guessing, with performance improving as the value approaches 1. Conversely, an AUC below 0.5 suggests that the classifier performs worse than random guessing; however, in such cases, inverting the prediction decisions can result in performance better than chance.

\end{appendix}



%

\end{document}